\documentclass[twocolumn,prl,aps,showpacs]{revtex4}

\usepackage{bm}
\usepackage{mathrsfs}
\usepackage{amsmath}
\usepackage{amssymb}
\usepackage{graphicx}
\usepackage{amsfonts}
\usepackage{amsthm}
\usepackage{color}
\usepackage{dcolumn}
\usepackage{txfonts}

\begin{document}

\title{Controllable effects of quantum fluctuations on spin free-induction decay at room temperature}
\author{Xin-Yu Pan}
\thanks{Email: xypan@aphy.iphy.ac.cn}
\author{Gang-Qin Liu}
\author{Dong-Qi Liu}
\affiliation{Beijing National Laboratory for Condensed Matter Physics and Institute of Physics,
Chinese Academy of Sciences, Beijing 100190, China}
\author{Zhan-Feng Jiang}
\author{Nan Zhao}
\thanks{Present address: 3rd Physics Institute and Research Center SCoPE,
University of Stuttgart, 70569 Stuttgart, Germany}
\author{Ren-Bao Liu}
\thanks{Email: rbliu@phy.cuhk.edu.hk}
\affiliation{Department of Physics and Center for Quantum Coherence, The
Chinese University of Hong Hong, Shatin, New Territories, Hong Kong, China}

\begin{abstract}
Fluctuations of local fields cause decoherence of quantum objects.
It is generally believed that at high temperatures, thermal noises
are much stronger than quantum fluctuations unless the thermal effects
are suppressed by certain techniques such as spin echo.
Here we report the discovery of strong quantum-fluctuation effects
of nuclear spin baths on free-induction decay of single electron spins
in solids at room temperature. We find that the competition
between the quantum and thermal fluctuations is controllable
by an external magnetic field.
These findings are based on Ramsey interference measurement of
single nitrogen-vacancy center spins in diamond and numerical
simulation of the decoherence, which are in excellent agreement.
\end{abstract}


\maketitle

Quantum systems lose their coherence when subjected to fluctuations of the local fields ($b$).
Such decoherence phenomena are a fundamental effect in quantum physics~\cite{Zurek_decoherence_PhysToday,Decoherence_classicalWorld,Schlosshauer_decoherence}
and a critical issue in quantum technologies~\cite{ClarkeSCqubit,LaddQC,Wrachtrup:2006,Chhildress06Science,MazeSensing,BalasubramanianMagnetometry,Zhao_magnetometry,Grinolds2011NP}.
The local field fluctuations can result from thermal distribution of the
bath states at finite temperature~\cite{Merkulov_decoherence_nuclei}, formulated as a density matrix
$\rho_{\text{E}}=\sum_J{p_J}|b_J\rangle\langle b_J|$ with probability $p_J$ for the local field
of a certain eigenvalue $b_J$. If the local field operator $b$ does not commute with the
total Hamiltonian of the bath $H_{\text{E}}$, a certain eigenstate $|b_J\rangle$ of
$b$ is not an eigenstate of the total Hamiltonian and will evolve to a superposition of
different eigenstates of $b$, causing quantum fluctuations of the local field~\cite{ZhaoADE}.
It is generally believed that at high-temperatures (as compared with transition
energies of the bath), the thermal fluctuations are much stronger than the
quantum fluctuations, though with certain control over the quantum systems,
such as spin-echo or dynamical decoupling control
in magnetic resonance spectroscopy~\cite{Hahn,DuUDD,CoryNVdd}, the decoherence effects of
thermal fluctuations can be largely suppressed.

In this Letter, we show that in the case of strong system-bath coupling (as compared with the
internal Hamiltonian of the bath), the quantum fluctuations can be comparable to
the thermal fluctuations, and induce notable effects even on free-induction
decay of the central spin coherence. The competition between the thermal and quantum fluctuations
can be controlled by an external magnetic field, indicated by crossover between Gaussian and non-Gaussian
decoherence accompanied by decoherence time variation.

The model system in this study is a nitrogen-vacancy center (NVC)
electron spin coupled to a bath of $^{13}$C nuclear spins
in diamond. This system has promising applications in quantum computing~\cite{Wrachtrup:2006,Chhildress06Science} and nano-magnetometry~\cite{MazeSensing,BalasubramanianMagnetometry,Zhao_magnetometry,Grinolds2011NP}.
The hyperfine interaction between the NVC spin and the bath spins is essentially dipolar and
therefore anisotropic. Due to the anisotropy of the interaction,  the hyperfine field
on a nuclear spin is in general not parallel or anti-parallel to the external magnetic field
and therefore the local Overhauser field $b$ (as a bath operator) does not commute with the Zeeman energy
of the bath. This induces strong quantum fluctuations, when the external field is not too strong
or too weak. The model system is representative of a large class of central spin decoherence
problems in which a central spin (such as associated with impurities or defects in solids)
has anisotropic dipolar interaction with bath spins~\cite{Dobrovtsiki09PRL}.

The NVC has a spin-1, which has a zero-field splitting $\Delta\approx 2.87$~GHz
between the states $|0\rangle$ and $|\pm 1\rangle$, quantized along the
$z$-axis (the nitrogen-vacancy axis). Since the NVC spin splitting is much greater
than the hyperfine interaction with the $^{13}$C spins, the center spin flip
due to the Overhauser field can be safely neglected~\cite{Chhildress06Science}. We only need to consider the
$z$-component of the local field fluctuation, $b_{z}=\sum_{j}{\mathbf
A}_{j}\cdot{\mathbf I}_{j},$ where ${\mathbf A}_{j}$ is the dipolar coupling
coefficients for the $j$th nuclear spin ${\mathbf I}_{j}$. The local field $b_z$ is a
quantum operator of the bath.
Within the timescales considered in this paper, the interaction between
the $^{13}$C nuclear spins, which has strength less than a few kHz~\cite{Maze08PRB,ZhaoNVdecoherence}, can be
neglected. The only internal Hamiltonian of the bath is the Zeeman energy
under an external magnetic field,
$H_{\text{E}}=\sum_j \gamma_{\text{C}}{\mathbf I}_j\cdot{\mathbf B}$, where $\gamma_{\text{C}}=6.73\times 10^7$~T$^{-1}$s$^{-1}$
is the gyromagnetic ratio of $^{13}$C.
To be specific, the magnetic field ${\mathbf B}$ is applied along the $z$ axis in this paper, but
the physics is essentially the same for field along other directions.
The Hamiltonian of the NVC spin and the bath can be written as~\cite{Maze08PRB,ZhaoNVdecoherence}
\begin{equation}
H=\Delta S_z^2+\left(\gamma_{\text{e}}B+b_z\right)S_z+H_{\text{E}},
\end{equation}
where $\gamma_{\text{e}}=1.76\times 10^{11}$~T$^{-1}$s$^{-1}$ is the electron gyromagnetic ratio,
and $S_z$ is the NVC spin operator along the $z$-axis.

\begin{figure}[t]
 \includegraphics[width=\linewidth]{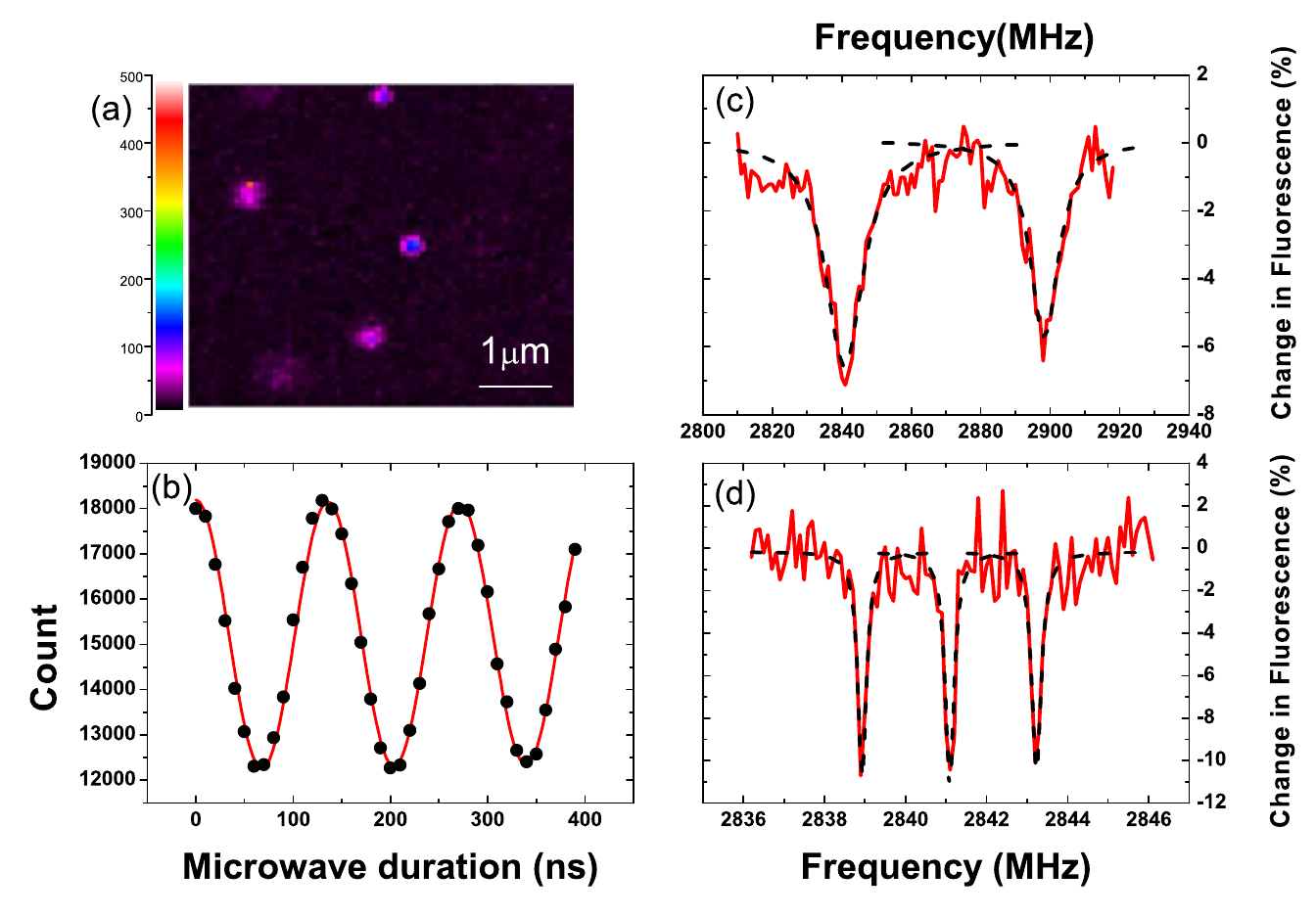}
  \caption{(Color online)
  (a) A fluorescence image of single NVC's in a type-IIa diamond.
  (b) Rabi oscillation of an NVC spin driven by a microwave pulse
  with the same strength as used in the Ramsey signal measurement.
  (c) Continuous-wave ODMR spectrum of an NVC spin,
     measured with a relatively strong microwave field
   (such that different lines due to different $^{14}$N nuclear spin states are
   not resolved). The two peaks (fitted with Lorentzian lineshapes in dashed lines)
   correspond to the transitions
   $|0\rangle\leftrightarrow|\pm 1\rangle$.
  (d)  Pulse ODMR spectrum near the
   $|0\rangle\leftrightarrow|- 1\rangle$ transition of an NVC spin,
   measured with a relatively weak microwave field
   (such that different lines due to different $^{14}$N nuclear spin states are
   resolved, fitted with Lorentzian lineshapes in dashed lines).
   The magnetic field is 10.3~Gauss in the measurement.}
  \label{Fig_setup}
\end{figure}

At room temperature, the nuclear spins are totally unpolarized. Thus the
bath can be described by a density matrix $\rho_{\text{E}}=2^{-N}I$,
with $N$ being the number of $^{13}$C included in the bath, and $I$ is
a unity matrix of dimension $2^N$. When the bath contains a large number
of nuclear spins (for example, $N>10$), the local Overhauser field
has a Gaussian distribution with width
\begin{equation}
\sigma=\left\langle b_z^2\right\rangle^{1/2}
=\frac{1}{2}\left(\sum_j A_j^2\right)^{1/2}.
\label{eq_sigma}
\end{equation}
This so-called inhomogeneous broadening would cause
a Gaussian decay of the NVC spin coherence,
$e^{-(t/T_2^*)^2}$ with the dephasing time $T_2^*=\sqrt{2}/\sigma$.

The quantum fluctuation of the local field $b_z$ arises from the
fact that in general $\left[b_z,H_{\text{E}}\right]\ne 0$,
especially when the nuclear Zeeman energy is comparable to the hyperfine coupling
$\gamma_{\text{C}}B\sim A_j$~\cite{ZhaoNVdecoherence}.
In the weak field case $\gamma_{\text{C}}B<< A_j$, the effect of the quantum
fluctuations would be negligible. In the strong field limit,
$\gamma_{\text{C}}B>> A_j$, the quantum fluctuation would also be suppressed,
since the nuclear spin flips due to the off-diagonal hyperfine interaction
(components of ${\mathbf A}_j$ perpendicular to the $z$-axis) would be
suppressed by the large Zeeman energy cost. In addition, the local field
fluctuation under a strong external field should
contain only the diagonal part, i.e., in Eq.~(\ref{eq_sigma}) for the
the inhomogeneous broadening, ${\mathbf A}_j$ should be replaced with
the $z$-component $A^z_j$. Therefore, we expect the dephasing time in the
strong field limit is longer than that in the weak field limit.
In the transition regime, the quantum fluctuation effect would be important,
and the dephasing would be in general non-Gaussian. Such features of
NVC center spin dephasing have been noticed previously in numerical simulations~\cite{ZhaoNVdecoherence}.

\begin{figure}[t]
 \includegraphics[width=\linewidth]{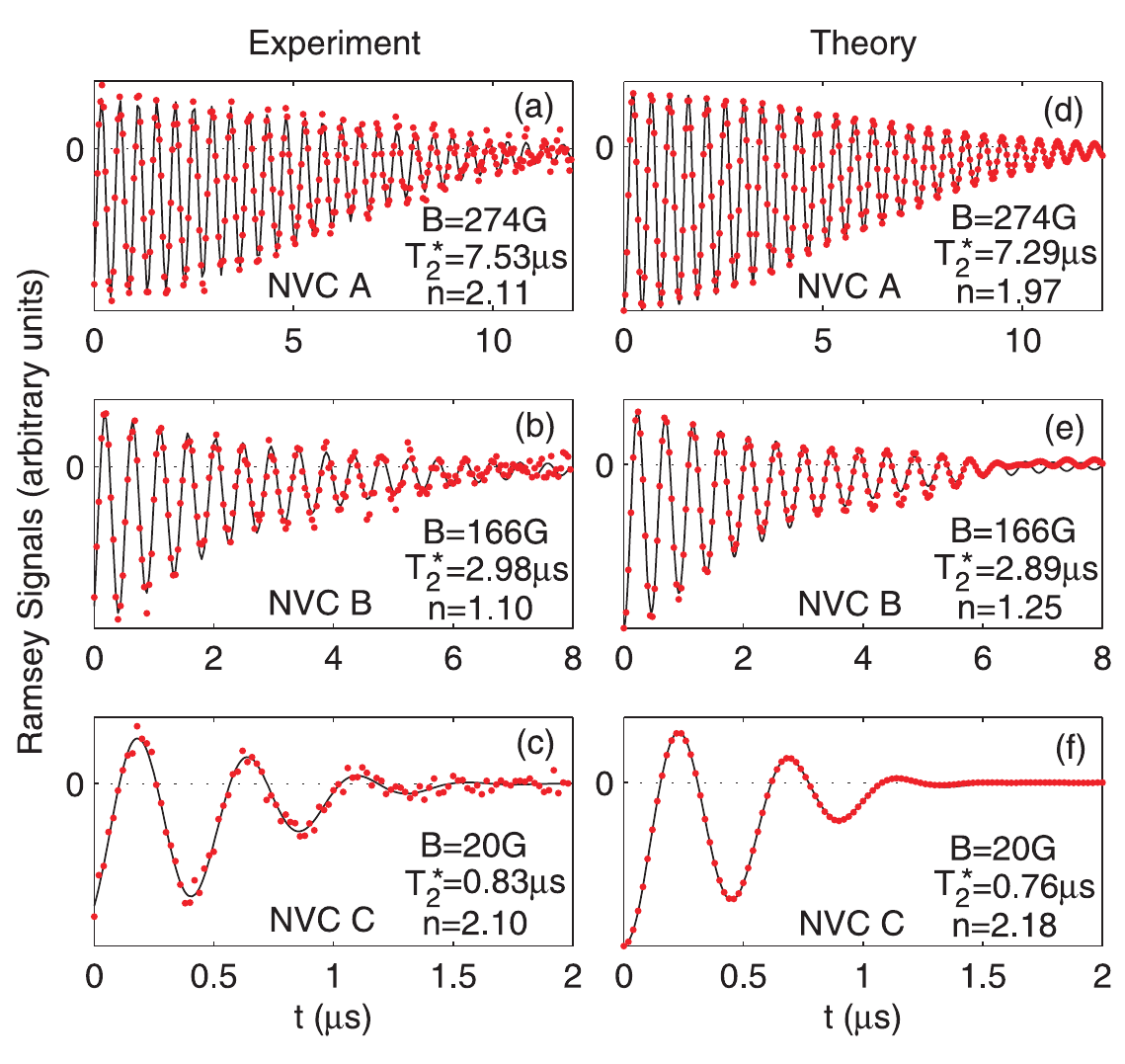}
  \caption{(Color online)
  (a), (b), and (c) in turn show three typical cases of experimentally measured Ramsey signals
  as functions of time for three NVC's A, B, and C under different magnetic fields.
  (d), (e), and (f) are numerical simulations corresponding to (a), (b), and (c) in turn.
  The red symbols are measured or calculated results, and the black lines  are fitting
  with Eq.~(\ref{Eq_fitting}).}
 \label{Fig_fitting}
\end{figure}

We use optically detected magnetic resonance (ODMR)~\cite{Gruber97Science} to measure the Ramsey interference
of single NVC spins in a high-purity type-IIa single-crystal diamond
(with nitrogen density $\ll1$~ppm)~\cite{Chhildress06Science}.
All the experiments are performed at room temperature.
Single NVC's in diamond are addressed by a home-built confocal microscope system
[see Fig.~\ref{Fig_setup}(a) for a typical fluorescence image of the single NVC's].
An external magnetic field is applied along the $z$-axis. The field
strength is tunable from 0 to 305~Gauss.
Under a weak field [10.3~Gauss as shown in Fig.~\ref{Fig_setup}(c)],
the two NVC spin transitions $|0\rangle\leftrightarrow|\pm 1\rangle$
are well resolved in spectrum. Furthermore, due to the hyperfine coupling
to the $^{14}$N nuclear spin, each NVC spin transition is split into three lines
corresponding to the three states of the $^{14}$N nuclear spin-1~\cite{Chhildress06Science},
which are resolved by weak-pulse ODMR measurement [see Fig.~\ref{Fig_setup} (d)
for the $|0\rangle\leftrightarrow|-1\rangle$ transition].
Fig.~\ref{Fig_setup}(b) shows the high-fidelity
rotation of the NVC spin under a microwave pulse of different durations.
The Ramsey interference measurement scheme is as follows:
The single NVC spin is first initialized to the state $|0\rangle$ by
optical pumping with a 532~nm laser pulse of 3.5~$\mu$s duration.
Then a $\pi/2$ microwave pulse excites the NVC spin to the superposition state
$\left(|0\rangle+|-1\rangle\right)/\sqrt{2}$.
The pulse is tuned resonant with the central line
(corresponding to the $^{14}$N spin state $|0\rangle_{^{14}\text{N}}$)
of the $|0\rangle\leftrightarrow|-1\rangle$ transition for each magnetic field.
The pulse duration [34~ns, corresponding to $\pi/2$ rotation in Fig.~\ref{Fig_setup}(b)]
is chosen as long as not to excite the $|0\rangle\leftrightarrow|+1\rangle$ transition
and as short as to spectrally cover all the three hyperfine lines corresponding to different $^{14}$N nuclear spin states.
After the first microwave pulse, the spin is left to freely precess about the
magnetic field with dephasing. After a delay time $t$, a second $\pi/2$ microwave pulse
is applied to convert the spin coherence to population in the state $|-1\rangle$.
The fluorescence of the NVC, which is about 30\% weaker when the spin is in $|-\rangle$
than it is when the spin is in $|0\rangle$,
is detected by photon counting under illumination of a 532~nm laser of 0.35~$\mu$s duration.
Each measurement (for a certain $B$ field and delay time $t$) is typically repeated $0.4\sim 1$ million
times to accumulate sufficient signal-to-noise ratio.

Typical Ramsey interference signals of single NVC spins are shown in Fig.~\ref{Fig_fitting} (a-c).
The oscillation is due to the beating between different transition lines corresponding to
the three $^{14}$N spin states~\cite{Chhildress06Science}. As shown in Fig.~\ref{Fig_fitting} (a-c),
the spin coherence represented by the fluorescence change
as a function of time, after subtraction of the background photon counting,
is well fitted with the formula
\begin{equation}
S=C e^{-\left(t/T_2^*\right)^n}\left[\frac{1}{3}+\frac{2}{3}\cos\left(A_{^{14}\text{N}} t+\phi\right)\right],
\label{Eq_fitting}
\end{equation}
in which $A_{^{14}\text{N}}$ is the hyperfine coupling constant to the
$^{14}$N nuclear spin,  $T_2^*$ gives the spin dephasing time,
and the exponential index $n$ characterizes the non-Gaussian
nature of the dephasing ($n=2$ corresponding to the Gaussian dephasing case).

\begin{figure}[t]
 \includegraphics[width=\linewidth]{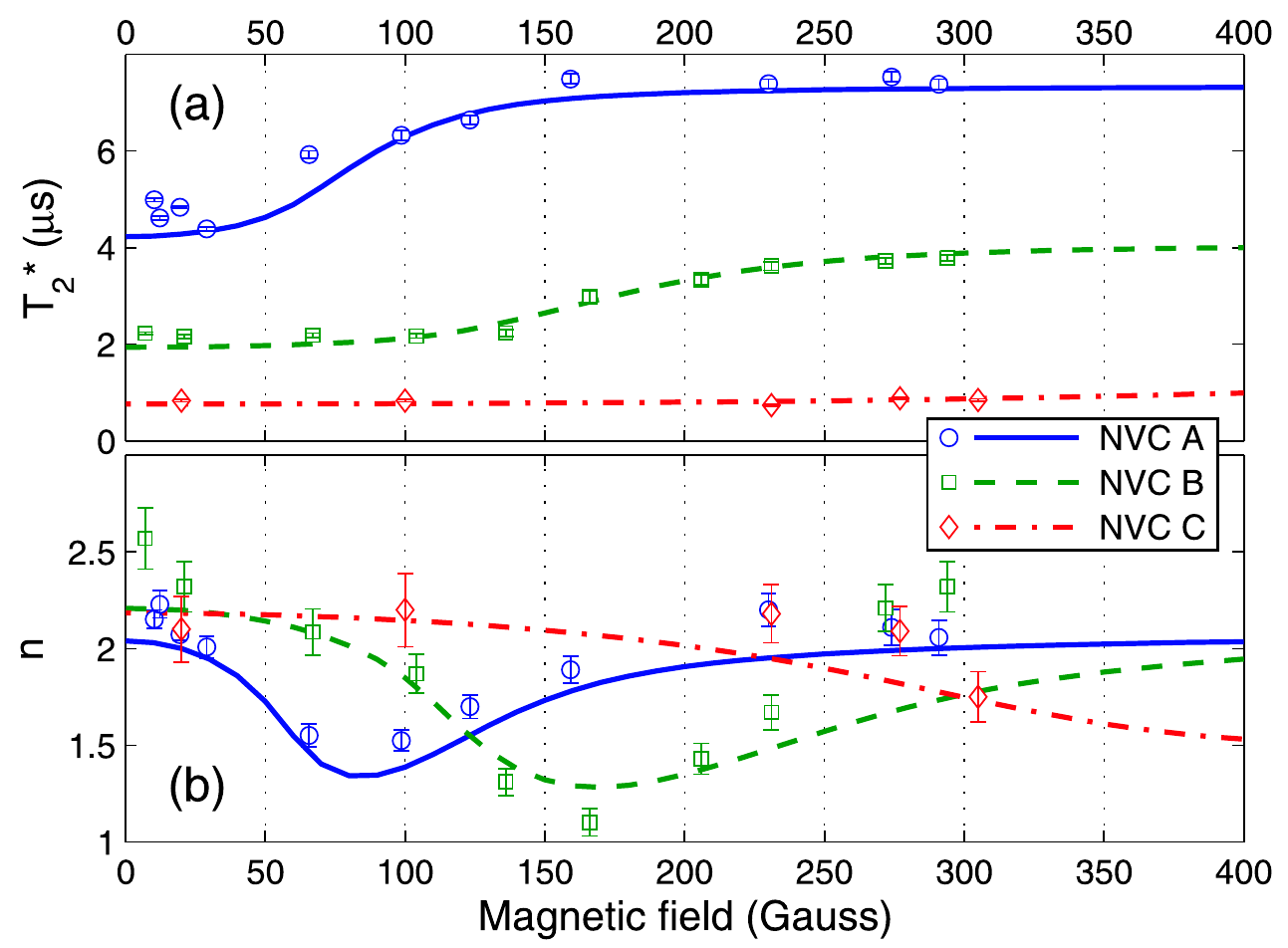}
  \caption{(Color online) Dependence on the magnetic field strength of
  (a) the dephasing time $T_2^*$ and (b) the exponential decay index $n$
  for three NVC's (A, B, and C) measured in experiments (circle, square, and diamond
  symbols with error bars),
  compared with the numerical simulations (solid, dashed, and dash-dotted lines).
  }
 \label{Fig_Compare}
\end{figure}

Figure~\ref{Fig_Compare} shows the spin dephasing time $T_2^*$ and the exponential decay index $n$ as functions of the
external magnetic field strength for three different NVC's (labeled A, B and C). The increasing of the dephasing time
with the magnetic field strength and the non-Gaussian decay associated with the dephasing time rising demonstrate
the competition between the thermal fluctuations of the local fields and the quantum fluctuations.
Since the $^{13}$C atoms (with abundance of 1.1\%) are randomly located around the NVC's,
the dephasing time $T_2^*$ presents a random distribution depending on the $^{13}$C position
configurations~\cite{ZhaoNVdecoherence}. An NVC with longer dephasing time should have $^{13}$C atoms located
farther away from the center with weaker hyperfine interaction (as the hyperfine interaction
is dipolar and decays rapidly with distance from the center).
Therefore, we expect that the quantum fluctuations for NVC's with longer dephasing times
start to take effect at lower magnetic field. This is indeed confirmed by the three sets of data
representing NVC's with long, intermediate, and short dephasing time (NVC A, B and C in turn).

To further confirm the physical picture of the quantum-thermal fluctuation crossover,
we carry out numerical simulations of the Ramsey signals with no fitting parameters.
Since the positions of the $^{13}$C atoms are not determined and the dephasing time
depends on the positions of the nuclear spins, we randomly choose the spatial configurations
such that the dephasing times at zero field are close to the
experimental values at the lowest field. The simulation is done with only single nuclear spins dynamics taken into account
(the interactions between nuclear spins are neglected since they are not relevant in the timescales
considered in this paper), which is an exactly solvable problem.
The Ramsey signal is given by~\cite{Maze08PRB,ZhaoNVdecoherence}
\begin{equation}
S
= \sum_{m=0,\pm 1} e^{im A_{^{14}\text{N}} t}
\prod_{n=1}^N \text{Tr}\left[
e^{i\gamma_{\text{C}}BI_j^zt}
e^{i {\mathbf A}_j\cdot{\mathbf I}_j t-i\gamma_{\text{C}}B I_j^zt}
\right].
\end{equation}
 As shown in Fig.~\ref{Fig_fitting} (d-f),
the calculated results are well fitted with Eq.~(\ref{Eq_fitting}).
In the simulations, the nearest 500 nuclear spins are included ($N=500$), which
produces well converged results. The dephasing time
and the exponential decay index obtained from the numerical results
are in excellent agreement with the experimental data (see Fig.~\ref{Fig_Compare}).

\begin{figure}[t]
 \includegraphics[width=\linewidth]{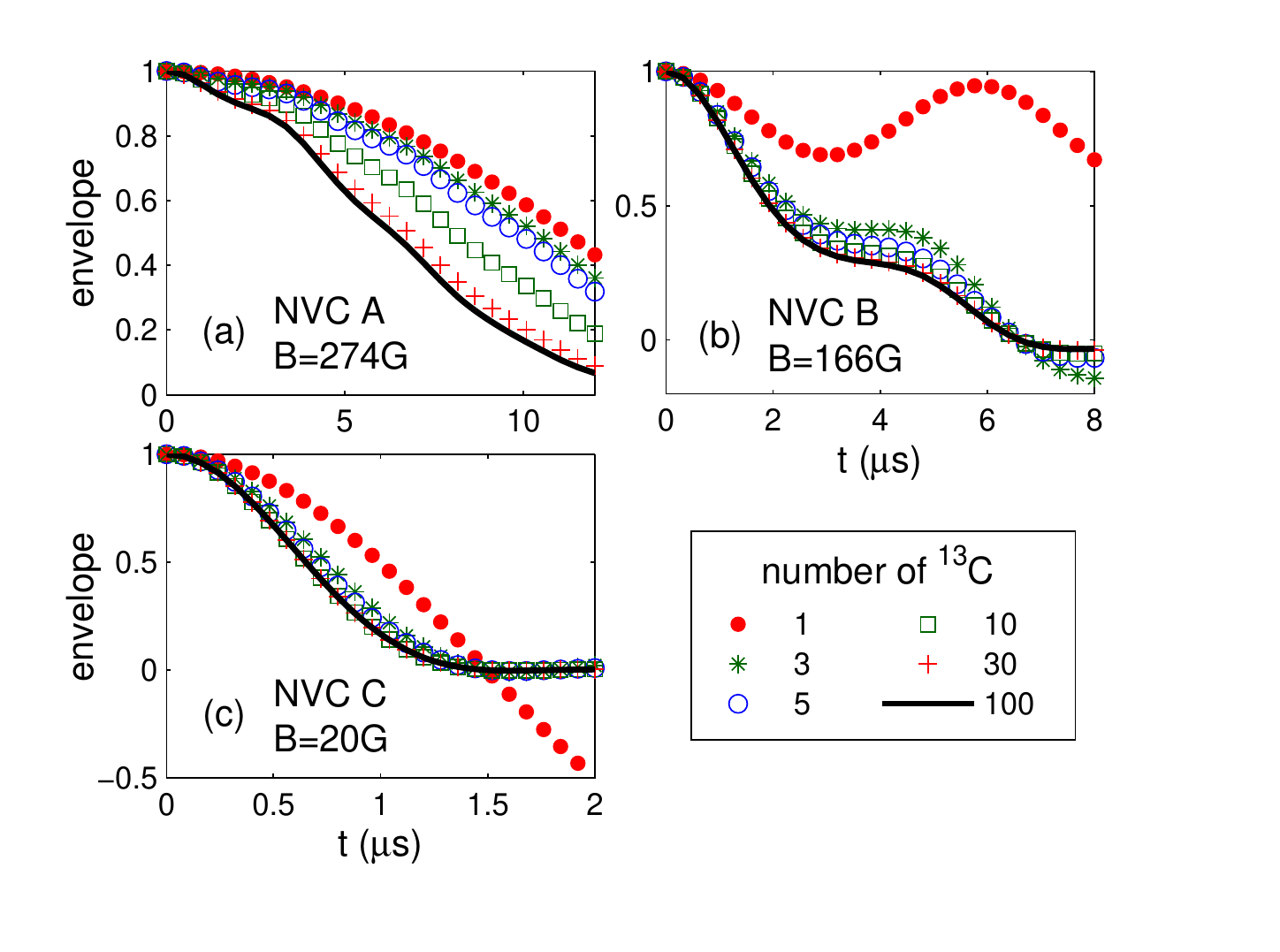}
  \caption{(Color online) Decay envelopes of the calculated Ramsey signals for
  various numbers of nearest $^{13}$C nuclear spins included in the bath, shown as filled circles,
  stars, open circles, open squares, crosses, and solid lines for $N=1$, 3, 5, 10, 30, and 100 in turn.
  (a), (b), and (c) are calculated under the same conditions as in
  Fig.~\ref{Fig_fitting} (d), (e), and (f) in turn.
  }
 \label{Fig_bathsize}
\end{figure}

Figure~\ref{Fig_bathsize} shows the contributions of nuclear spins at different distances
to the NVC spin dephasing. The nearest few $^{13}$C nuclear spins
already contribute the major part of the local field fluctuations.
A close examination of the $^{13}$C positions in different configurations
reveals that the average hyperfine coupling constants for
the nearest 10, 5, and 3 nuclear spins (which contribute
the major part of the dephasing) for NVC A, B, and C
are $\bar{A}\approx 0.16$, 0.51, and 1.7~$\mu$s$^{-1}$ in turn.
Correspondingly, the quantum fluctuations should
start to take effect at magnetic field strength $B\sim \bar{A}/\gamma_{\text{C}}\approx 24$,
76, and 260~Gauss for NVC A, B, and C in turn.
This is indeed consistent with the experimental observation shown in Fig.~\ref{Fig_Compare}.

In conclusion, we demonstrate that even at room temperature (which can be
regarded as infinite for the nuclear spin baths considered here)
and in free-induction decay of spin decoherence,
the quantum fluctuations of local fields can be as strong as the thermal fluctuations
in a mesoscopic spin bath with anisotropic interaction with the central spin.
The contribution of the quantum fluctuations can be tuned by an external magnetic field.
In addition to revealing an aspect of the quantum nature of nuclear spin baths,
the effect can be used to identify optimal physical systems and parameter ranges
for quantum control over a few nuclear spins via a central electron spin.
Such control is relevant to quantum computing and nano-magnetometry~\cite{Wrachtrup:2006,Chhildress06Science,MazeSensing,BalasubramanianMagnetometry,Zhao_magnetometry,Grinolds2011NP}.

\begin{acknowledgments}
This work was supported by National Basic Research Program of China (973 Program
project No. 2009CB929103), the NSFC Grants 10974251 and 11028510,
Hong Kong RGC/GRF CUHK402208 and CUHK402410, and CUHK Focused Investments Scheme.
\end{acknowledgments}


\end{document}